\documentclass{egpubl}
\usepackage{pg2026s}

\makeatletter
\let\ps@titlepage@orig\ps@titlepage
\def\ps@titlepage{\ps@titlepage@orig \def\@oddhead{}\let\@evenhead\@oddhead}
\makeatother
\copyrightTextTitPag{
    This is the pre-peer reviewed version of the following article:
    \textsc{Radovanovic V., Gupta V., Gruson A., Hua B.-S.}:
    Physically Based Rendering in the Latent Space.
    \textit{Computer Graphics Forum} (2026),
    which will be published in final form at
    \url{https://doi.org/10.1111/cgf.70633}.
    This article may be used for non-commercial purposes in accordance with Wiley Terms and Conditions for Use of Self-Archived Versions.
}
\copyrightTextRunPag{}

\usepackage[T1]{fontenc}
\usepackage{dfadobe}  

\usepackage{cite}  %
\BibtexOrBiblatex
\electronicVersion
\PrintedOrElectronic
\ifpdf \usepackage[pdftex]{graphicx} \pdfcompresslevel=9
\else \usepackage[dvips]{graphicx} \fi

\usepackage{egweblnk}

\usepackage{hyperref}
\usepackage{amsmath}
\usepackage{amssymb}
\usepackage{listings}
\usepackage[capitalize,nameinlink]{cleveref}
\usepackage{xspace}
\usepackage{tabularx}
\usepackage{mathtools}
\usepackage{booktabs}
\usepackage{multirow}
\usepackage{enumitem}
\usepackage{graphicx}   %
\usepackage{makecell}   %
\usepackage{siunitx} %
\usepackage{microtype}
\usepackage{xcolor}

\definecolor{myblue}{HTML}{4C72b0}
\definecolor{myorange}{HTML}{DD8452}
\definecolor{myyellow}{HTML}{CC9900}
\definecolor{mygreen}{HTML}{55A868}
\definecolor{myred}{HTML}{C44E52}
\definecolor{mypurple}{HTML}{8172B3}
\definecolor{success}{HTML}{55A868}
\definecolor{failure}{HTML}{C44E52}

\newcommand{\enc}{\mathcal{E}}
\newcommand{\dec}{\mathcal{D}}

\newcommand{\xp}{\mathbf{x}}
\newcommand{\yp}{\mathbf{y}}
\newcommand{\xpath}{\bar{\mathbf{x}}}
\newcommand{\pixelfilter}{r}
\newcommand{\bsdf}{f_r}
\newcommand{\contribfunc}{f}
\newcommand{\T}{T}
\newcommand{\pathspace}{\mathcal{P}}
\newcommand{\sceneparams}{\pi}
\newcommand{\sceneparamsopt}{\hat{\sceneparams}}
\newcommand{\MSE}{\mathcal{L}_\mathrm{MSE}}
\newcommand{\SSIM}{\mathcal{L}_\mathrm{SSIM}}
\newcommand{\LPIPS}{\mathcal{L}_\mathrm{LPIPS}}

\title{Physically Based Rendering in the Latent Space}

\author[V. Radovanovic et al.]{
    \parbox{\textwidth}{\centering
        Vuk Radovanovic$^1$ \orcid{0009-0008-9271-9149}
        Vishesh Gupta$^1$ \orcid{0009-0005-3201-9827}
        Adrien Gruson$^2$ \orcid{0000-0003-0773-2725}
        Binh-Son Hua$^1$ \orcid{0000-0002-5706-8634}
    }
    \\
    \parbox{\textwidth}{\centering
        $^1$Trinity College Dublin \\
        $^2$École de Technologie Supérieure (ÉTS)
    }
}

\begin{document}

\teaser{
  \centering
  \includegraphics[width=\linewidth]{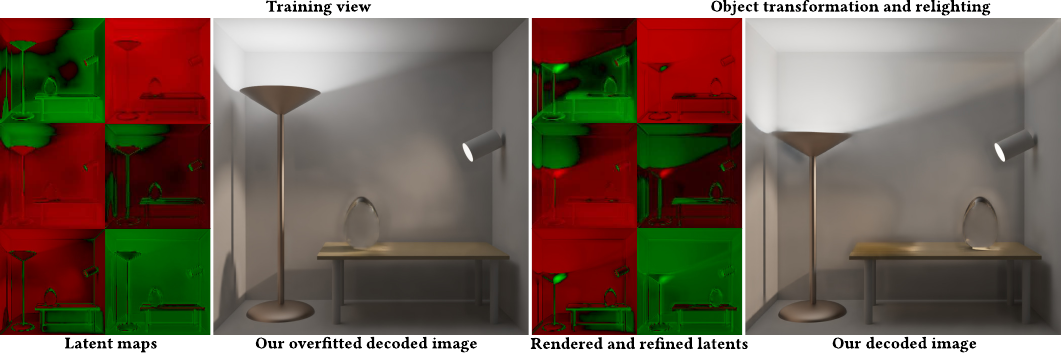}
  \caption{We introduce latent rendering, an image synthesis paradigm that leverages light transport simulation to directly render latent representations of autoencoders. The rendered latents can be subsequently decoded to output final color rendering. This paradigm allows us to perform physically based rendering on 3D scenes to output latents for scene editing applications. We train the renderer using a single image (left), and show its generalized rendering to a scene variant (right, with lamp shortened and glass egg shifted). Latent maps are color-coded, with green for positive and red for negative values, with colors fading to black as values approach zero.
  }
  \label{fig:teaser}
}

\maketitle
\begin{abstract}
  Image diffusion models have shown impressive image generation capabilities but are often hard to control, in contrast to classical computer graphics pipelines such as physically based rendering. However, we observe that there is a bridge between light transport phenomena and the distribution of latent space values produced by such models. Thus, we introduce physically based rendering in the feature space learned by the variational autoencoders in generative models, enabling light transport simulation in the latent space. This allows us to leverage physically based rendering techniques to output latent maps for physically guided content generation. We propose modifications to the rendering equation, which, when paired with a differentiable renderer, can yield an optimal set of scene parameters that require only minimal refinement to accurately render into the pretrained latent space. We train our method on a single rendered image, and then demonstrate the generalization of the method to scene geometry changes, lighting changes, and camera view changes.

\ccsdesc[500]{Computing methodologies~Ray tracing}
\ccsdesc[300]{Computing methodologies~Machine learning}

\printccsdesc   
\end{abstract}  

\section{Introduction}
The history of computer graphics is marked by a rich family of techniques aiming to solve the longstanding challenge of realistic image synthesis.
Notable approaches include rasterization-based methods for real-time graphics and computer games, along with global illumination methods for high-end animation, movie production, and architectural visualizations. Among these methods, Monte Carlo rendering has been popular across applications because it can simulate physically based light transport effects by numerically solving the rendering integral~\cite{Kajiya:1986:Rendering}. 
Several rendering algorithms can be used to estimate this integral, namely path tracing~\cite{Kajiya:1986:Rendering}, bidirectional path tracing~\cite{veach1995bidirectional,lafortune1993bidir}, photon mapping~\cite{jensen1996photon,hachisuka2008ppm}, and Metropolis light transport~\cite{veach1997metropolis}. 

The recent breakthrough of generative modeling in artificial intelligence, i.e., generative AI, has led to radical changes in visual computing, particularly in how visual content is created in computer graphics and computer vision. 
Particularly, diffusion models~\cite{sohl2015thermo,song2019generative,ho2020ddpm} demonstrate great promise in solving image synthesis in computer graphics from a new perspective: by generating photorealistic images and videos using natural language and visual descriptions~\cite{rombach2022high,xie2024sana,bfl2025representation,wan2025wan}. 
Such generative models are purely data driven, bypassing the traditional graphics pipeline for visual synthesis.

A key challenge in the latest wave of generative AI models is that these models are not grounded in the laws of physics. Most generative AI models learn features from large-scale datasets and synthesize data by learning and subsequently sampling from the training data distributions. Physical plausibility in visual content generation is mainly driven by data, leaving physical accuracy and correctness largely underconstrained. 
While computer graphics has numerous techniques targeting physical correctness, generative models lack a principled framework rooted in a physical basis.

This work is our first step toward applying physically based rendering constraints to generative models. We derive a new rendering paradigm that aims to integrate the rigor of light transport simulation with the expressiveness and scalability of visual content generation.
We draw inspirations from state-of-the-art image diffusion models, which operate with three main components: a variational autoencoder (VAE), a denoising network, and a prompt conditioning module.
Among these, the VAE defines an encoder and decoder that transforms a color image into a latent space that is highly compressed, but remains structured and interpretable; i.e., the latent maps of an image bear high similarity to the geometric and shading structures that appear in the full, uncompressed image.
Motivated by this design, we propose \emph{latent rendering}, a governing approach that connects physically based rendering and generative models via the latent space. 
The goal is to repurpose light transport simulation to operate in this latent space and output latent maps, which can be subsequently reconstructed to color images. 
By learning to render in the latent space, we aim to understand its connection to traditional RGB rendering, facilitating research toward physically-grounded generative rendering. 

Compared to traditional rendering in the RGB color space, latent rendering offers the following benefits. 
First, latent space is significantly lower-dimensional than RGB space: e.g., a $1024 \times 1024 \times 3$ image could be encoded to a $128 \times 128 \times 16$ latent map, a $12 \times$ reduction that makes rendering operations more compact in the latent space. 
Second, by rendering into the latent space, we can generalize to render larger latent maps, e.g., $256 \times 256 \times 16$, enabling even higher resolution RGB decoding, e.g., to $2048 \times 2048$. 
For downstream applications, we demonstrate that it is possible to more efficiently perform language-guided editing on scene representations using score distillation sampling methods~\cite{poole2023dreamfusion}. In such applications, latent rendering allows us to circumvent the image encoding step that would otherwise be necessary for an RGB renderer to interface with a latent diffusion model.  In our case, gradients backpropagate to directly update scene parameters without passing through the encoder.

To bring this idea to life, we propose the very first latent rendering algorithm. Our formulation is derived from our analysis of latent representations obtained from encoding a physically based rendering. This analysis motivates us to adapt light transport simulation to reproduce the visual effects perceivable in the latent space, leading to our latent rendering formula. This latent rendering process can be optimized based on differentiable rendering and refined using a neural network, making the rendered latents accurately match ground truth representations using only a single RGB rendering for training. The latent representations can then be edited via 3D scene variants under camera movement, light source changes, or object motion, to produce updated latents that can be reconstructed to color rendering. 
To summarize, our main contributions are:
\begin{itemize}[leftmargin=*]
    \item A thorough analysis of the latent space of a pretrained autoencoder through the lens of physically based light transport;
    \item A latent rendering formulation including a modified rendering equation and a neural refiner which depart from its original physical basis, but provide adequate expressibility to mimic generative latent representations;
    \item A pipeline allowing to freely edit camera poses, light configuration, and object position directly in latent space for physically based scene-aware image editing. 
\end{itemize}

Our code is available at \url{https://github.com/trinity-graphics/latent-rendering}.

\section{Related Work}

\subsection{Forward and inverse physically-based rendering} 
In computer graphics, light transport simulation has been the core of physically based rendering over the past decade~\cite{pharr2023pbrt}. 
The rendering equation~\cite{Kajiya:1986:Rendering} is the central formulation that dictates how light emitted from sources interacts with 3D objects through their associated bidirectional scattering distribution functions (BSDFs) and is ultimately captured by a virtual camera to produce a photorealistic image. 
Different solutions to this rendering equation exists, such as path tracing~\cite{Kajiya:1986:Rendering}, photon mapping~\cite{jensen1996photon,hachisuka2008ppm}, and Markov Chain Monte Carlo rendering~\cite{veach1997metropolis}. 
Among these, path tracing~\cite{Kajiya:1986:Rendering} remains the de facto standard used in industry, with several improvements such as path guiding~\cite{herholz2025pathguiding}, path reuse~\cite{Bekaert2002PathReuse}, and reservoir sampling~\cite{wyman2023restir}.
A key benefit of physically-based rendering is that this rendering pipeline is fully explainable and controllable, offering fine-grained and precise control of image appearance through explicit scene editing using interactive authoring tools~\cite{blender}. 

Light transport simulation can be repurposed to operate in spaces other than the original RGB space.  
Gradient-domain rendering~\cite{Lehtinen2013GradientMLT,kettunen2015gdpt,Bauszat2017GradientPathReuse,Hua2019GradientSurvey} simulates light transport to estimate image gradients, which are used subsequently for RGB reconstruction~\cite{yan2025generalizedrecon}, taking advantage of the spatial smoothness of rendered images.
In gradient-domain rendering, these gradients remain linear to their underlying pixel values. 
Our latent rendering instead operates in a non-linear compressed data representation learned by an encoder-decoder pair.

Light transport simulation could also be made differentiable for inverse rendering tasks~\cite{Li:2018:DMC,NimierDavidVicini2019Mitsuba2,Zhang2020PathSpaceDiff,zeng2025surveydr}, e.g., for optimizing scene parameters to match a target image through gradient backpropagation.
By applying Reynolds' transport theorem~\shortcite{reynolds1903papers}, the differentiation of the rendering equation yields an interior and an exterior term~\cite{zeng2025surveydr}. The exterior term models signal discontinuities such as visibility changes, and must be handled when optimizing scene geometry, typically requiring specialized techniques~\cite{Zhang2023Projective,Xu2024MCMCBoundary,Soroka2025SilhouetteQuadric}. The interior term describes how scene parameters, such as BSDFs or emitter profiles, influence the rendered image. These interior derivatives can be efficiently evaluated using backward radiance formulations~\cite{NimierDavid2020Radiative}, e.g., path-replay backpropagation~\cite{Vicini2021PathReplay}. In this work, we built on top of differentiable rendering to redirect light transport simulation to operate in the latent space, exclusively focusing on the interior component for optimizing emitter and BSDF parameters.

\subsection{Generative models} In contrast to physically based rendering, generative models offer a new paradigm for synthesizing photorealistic images by learning from data, e.g., using diffusion models~\cite{sohl2015thermo,ho2020ddpm,song2019generative}. 
Latent diffusion models~\cite{rombach2022high,xie2024sana} perform the diffusion process in highly compressed but interpretable latent spaces, enabling large-scale training and high-resolution image generation. 
Video diffusion models~\cite{yang2024cogvideox,wan2025wan} are recent advances that extend image diffusion models for generating animated scenes. 
While the latent spaces of these generative models are highly expressive, controlling the outputs of diffusion models is a challenging task. 
General methods such as low-rank adaptation~\cite{hu2022lora} and ControlNet~\cite{zhang2023controlnet} adjust a subset of parameters of the diffusion models to induce specific visual styles, with optional extra inputs such as depth maps or user-provided strokes to control the generation process. 
The denoising diffusion process could also be intercepted with visual, semantic, or user guidance at early timesteps to provide high-level control for image editing~\cite{meng2022sdedit}. 
Alternatively, specific solutions are required for particular tasks such as intrinsic decompositions~\cite{lyu2025intrinsicedit,he2025neurallightrig}, material generation~\cite{kocsis2025intrinsix}, relighting~\cite{zeng2024rgb2x,bharadwaj2025genlit}, or novel view synthesis~\cite{liu2023zero1to3,zhou2025seva}. 
However, these approaches do not offer sufficiently precise or consistent control over camera parameters, lighting conditions, and object placement simultaneously.

\subsection{Hybrid rendering}
There exist ongoing efforts in combining generative models with traditional rendering concepts to model physically based effects and improve controls over image outputs. 
For example, transformer-based models have been proposed to directly predict approximate solutions to the rendering equation~\cite{zeng2025renderformer}, but they remain limited to relatively small scenes. 
Video diffusion models can be tuned to output material and environment maps, enabling the generation of material and lighting effects in videos without light transport simulation~\cite{liang2025diffusionrenderer}.
Neural renderers could be used to refine physically based rendering results, enabling tasks such as scene  relighting from a single image~\cite{careaga2025PBRRelighting} when combined with monocular geometry estimation~\cite{wang2025moge}.  %

A few previous works exploit the latent space for feature rendering. 
Latent-NeRF~\cite{metzer2023latentnerf} performs text-to-3D generation by directing a neural radiance field to output latent features compatible to a pretrained diffusion model, as guided by a score distillation objective. 
Splatent~\cite{hirschorn2025splatent} reconstructs 3D Gaussian Splats in the latent space of a VAE, and leverages a one-step diffusion process to improve the quality of the latents, thereby enhancing novel view rendering.  
Our work shares the high-level idea of rendering to the latent space as we aim to use the latent space to establish a synergy between latent diffusion models and physically based light transport simulation. 
As our method is built on physically based rendering, relighting is naturally supported and scene editing is generally more intuitive and controllable.

\section{Background}

\subsection{Physically based rendering} We are interested in solving the light transport equation~\cite{veach1998robust}, which, a light path $\xpath = {\xp_0, \ldots, \xp_k}$ consisting of $k$ vertices, contributes to pixel $i$ by evaluating the following integral over the path space $\pathspace$: 
\begin{align} \label{eq:path-space}
I_i(\sceneparams) = \int_{\pathspace} \contribfunc_i(\xpath, \sceneparams) d\xpath
= \int_{\pathspace} \pixelfilter_i(\xpath) \T(\xpath, \sceneparams) L_e(\xp_{k-1} \leftarrow \xp_k, \sceneparams) d\xpath .
\end{align}
Here, $\pixelfilter_i(\xpath)$ denotes the pixel filter, $\T(\xpath, \sceneparams)$ the path throughput, and $L_e(\xp_{k-1} \leftarrow \xp_k, \sceneparams)$ the emitted radiance at the light source. 
The throughput is defined as $\T(\xpath) = \prod_{j=1}^{k-1} \bsdf(\xp_j,\sceneparams)\cos\theta_j$, where $\bsdf$ is the bidirectional scattering distribution function (BSDF) evaluated at each surface interaction, and $\theta_j$ is the angle between the incident direction at vertex $j$ and the surface normal. 
The integral depends on scene parameters $\sceneparams$, such as BSDF parameters and emitter intensity values. 
The integral in \cref{eq:path-space} can be solved using Monte Carlo estimation. 
By default, the integral is evaluated in standard color spaces such as the RGB and produces non-negative estimates that respects physical constraints imposed by light transport.

\subsection{Differentiable rendering}
Physically based rendering can be used in an optimization framework to determine optimal scene parameters $\sceneparamsopt$ so that the rendering matches a target image $I_\mathrm{gt}$: 
\begin{align}
\sceneparamsopt = \arg \min_{\sceneparams}\; \mathcal{L}(I(\sceneparams), I_\mathrm{gt}),
\end{align}
where $\mathcal{L}$ is a differentiable objective function such as the mean squared error (MSE). Gradient-based optimizers, such as Adam~\cite{kingma2014adam}, can be used to solve this optimization problem, but they require access to how changes in the scene parameters affect the rendered image. By the chain rule, this requires computing the gradient of each pixel intensity with respect to the scene parameters:
\begin{align}
\partial_{\sceneparams} I_i(\sceneparams)
= \partial_{\sceneparams} \int_{\pathspace} f_i(\xpath, \sceneparams) d\xpath
= \int_{\pathspace} \partial_{\sceneparams} \contribfunc_i(\xpath, \sceneparams) d\xpath .
\end{align}
Transferring the derivative inside the integral requires care when the scene parameters control discontinuities in the rendering process \cite{reynolds1903papers}. 
Several techniques for differentiable Monte Carlo rendering exists~\cite{Li:2018:DMC,Vicini2021PathReplay}, but in this work, we assume that our scene parameters vary continuously over the integration domain, and their derivatives could be efficiently estimated using path replay backpropagation~\cite{Vicini2021PathReplay}.

\subsection{Latent spaces}
Modern generative models~\cite{rombach2022high,bfl2025representation} are typically trained in two stages. First, an autoencoder is trained to learn a latent space that encodes data into latent variables and reconstructs the original data from these latent representations. Second, a denoising network is learned to map random noise to latent codes. 
In this work, we focus on the first stage, namely the autoencoder and its latent representations.

An autoencoder is composed of an encoder $\enc$ and decoder $\dec$. Given an image $x \in \mathbb{R}^{H \times W\times 3}$, the encoder produces a compressed latent representation $z =\enc(x) \in \mathbb{R}^{h \times w \times d}$ such that $h < H$, $w < W$, and $d > 3$ where $H,W,h,w$ are image and latent height and width, and $d$ is the number of latent channels, respectively. We can then decode this latent representation to reconstruct an image  $\tilde{x}=\dec(z) \in \mathbb{R}^{H \times W \times 3}$ similar to original image $x$. This autoencoder is trained to minimize the difference between the distribution of $x$ and $\tilde{x}$, following the principles of variational autocoders (VAEs)~\cite{kingma2013vae}.
From empirical designs of image diffusion models, typically we have $h = H / 8$, $w = W / 8$, $d=4$ for Stable Diffusion 2.1 and SDXL~\cite{rombach2022high}, and $d=16$ for Stable Diffusion 3.5 and FLUX~\cite{esser2024scaling,bfl2025representation}, resulting in latent compression ratio of 48 and 12, respectively.

\section{Latent Space Analysis}
\label{sec:latentanalysis}

\begin{figure}[t]
    \centering
    \includegraphics[width=\linewidth]{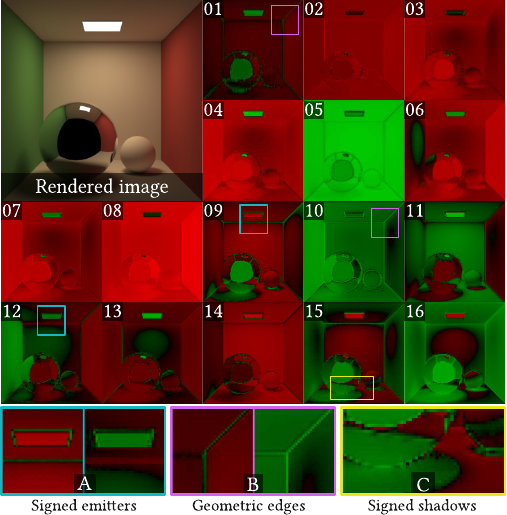}
    \vspace{-6mm}
    \caption{Latent maps of the Cornell Box rendering. The zoom-in patches demonstrate visual effects unique to latent representations, including (A) the signed emitter, (B) highlighted geometry edges, and (C) signed shadows, which cannot be rendered using traditional physically based rendering. 
    }
    \label{fig:latentcornel}
    \vspace{-4mm}
\end{figure}

This work aims to align rendering values produced by a physically based renderer with autoencoder latents, enabling the renderer to generate latent representations that can subsequently be decoded into color images. Let us first analyze these latents to identify the requirements for adapting a physically based renderer to the latent space.  
Here we focus on the latent space defined by the pretrained VAE in Stable Diffusion 3.5~\cite{esser2024scaling}, an open-weights text-to-image diffusion model for high-quality image generation. 
We use image resolution of $1024 \times 1024$, latent resolution $128 \times 128$ with $d = 16$.
We analyze the latents of the classic Cornell box as shown in~\cref{fig:latentcornel}. 
We observe the following phenomena that makes rendering in this latent space inherently different from RGB rendering:
\begin{enumerate}[leftmargin=*]
    \item In the latent space, emitted light is signed (\cref{fig:latentcornel}, A), in contrast to the strictly non-negative values assumed in physically based rendering.
    \item The latents encode scene structure with a strong emphasis on geometric edges and corners (\cref{fig:latentcornel}, B). In regions with extremely high-frequency content, latent values may become perceptually indistinguishable from noise.
    \item In occluded regions, latent values can exhibit both positive and
    negative energy, which can be an even higher intensity than surrounding unoccluded regions (\cref{fig:latentcornel}, C). This differs from physically based rendering, where occlusion from a light source can only reduce the amount of energy received by a surface.
\end{enumerate}

\begin{enumerate}[leftmargin=*]
    \item There is structural parity between the RGB image and the corresponding latent channels, in which different BSDFs and objects receive distinct latent values that mirror their RGB counterparts.
    \item Signed energy emitted in the latent space is reflected throughout the scene and is closely preserved on perfectly specular surfaces. 
    \item As the incident angle of light varies smoothly, the intensity of the reflected signed energy varies accordingly. This effect is more pronounced in some channels and less so in others, with certain channels exhibiting a flatter, more uniform appearance (e.g., \cref{fig:latentcornel}, channel 02). %
\end{enumerate}

In the literature, there exist some attempts in modeling the relationship between latents and RGB values directly, such as a 4x4 linear mapping in Stable Diffusion 1.4~\cite{Turner2022LatentToRGB}. 
However, such mapping is applied for general images and to each pixel independently. 
Our analysis indicates that we can take a further step by simulating light transport in the latent space. To achieve this, the rendering process must be adapted to freely express the range of latent values observed. 

\section{Method}
Our pilot study suggests that the latent representations might consist latent values producible by physically based rendering, and residual values that require additional modeling, e.g., high-frequency latents at shape boundaries. 
Based on these observations, we propose to disentangle the latents into two components: physically based latents and residual latents. 
We leverage light transport simulation to render the physically based latents, and use a neural refinement network to model the residual latents. 
This disentanglement allows us to separately address physically grounded effects and non-physically based phenomena present in the latents in a principled manner.

\begin{figure*}[t!]
    \centering
    \includegraphics[width=\linewidth]{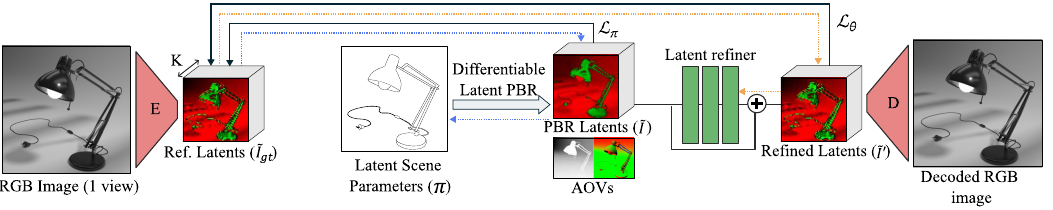}
    \caption{We train a differentiable renderer to output to the latent space defined by a variational autoencoder. We assume a 3D scene is available for our rendering task. Our training pipeline consists of taking a ground truth RGB rendering of the scene and encoding it into the latent space. We then optimize the 3D scene parameters to make its rendering reproduce this latent representation using our modified rendering equation (\cref{eq:latenteq}). The rendered latents are then refined using a per-scene refiner network. The final latents are decoded to reconstruct the RGB rendering. Once the pipeline is trained, it can be applied to render latents for variants of the same scene with edited scene settings. The {blue dotted} line represents backpropagation for scene parameter optimization. The {yellow dotted} line represents backpropagation for the refiner parameter training.}
    \label{fig:pipeline}
\end{figure*}

Our full pipeline is summarized in \Cref{fig:pipeline}. First, we apply differentiable rendering to train our physically based renderer such that it can output latent values by optimizing the scene parameters ($\sceneparams$) to match reference latent images ($\tilde{I}_\mathrm{gt}$), following our modified latent rendering formulation (\cref{subsec:latentrender}). We then predict the residual latents using a neural refinement network (\cref{subsec:neuralrefiner}), which learns to compensate for discrepancies between the rendered latents and the target latents, particularly over geometric structures such as edges and corners. Details of our pipeline are as follows.

\subsection{Latent rendering}
\label{subsec:latentrender}

As discussed in the latent analysis section (\cref{sec:latentanalysis}), the latent space includes both positive and negative values, whereas conventional physically based rendering produces only non-negative values. A naïve way to adapt a physically based renderer for signed rendering is to shift latent values into the non-negative domain by applying an offset and a scale on the latents, or by splitting the positive and negative latents and perform negation on the negatives. 
However, we find that both approaches are not effective and efficient to address the observations as outlined in \cref{sec:latentanalysis}. Particularly, shifting the values will result in appearance changes while splitting the latents requires doubling the number of latent channels. 
Reconstructing the modified latents is challenging as well as the decoder requires to de-normalize the latents to their vanilla domain before decoding. 
Instead of processing the latents heuristically, we therefore opt for a principled approach by modeling the latents using the rendering equation. 
Let us begin by introducing the concept of \textit{signed rendering} as follows. 

\subsubsection{Signed radiance and reflectance} 
In keeping with our goal of strong physically-based priors, we begin our latent rendering approach by modeling the latents' physically-aligned features: structure, shading, and global illumination. 
In traditional rendering, emitted light $L_e$ and the BSDF $\bsdf$ are constrained to be non-negative in order to ensure valid light path sampling. To accommodate signed latent values, we modify the traditional path tracing approach to support the propagation of both positive and negative energy throughout the scene.  First, we allow emitters to emanate signed light energy $\tilde{L}_e(\xp \leftarrow\yp,\sceneparams)$, so that $\tilde{L}_e$ can emit positive or negative radiance values.
We also modify BSDF evaluations to enable them to yield signed values.
This leads to the following signed light transport equation:
\begin{align} \label{eq:signedrender}
\tilde{I}_i^e(\sceneparams)=\int_\pathspace\pixelfilter_i(\xpath) 
\tilde{T}(\xpath,\sceneparams)\tilde{L}_e(\xp_{k-1}\leftarrow\xp_k, \pi)d\xpath \,,
\end{align}
where $\tilde{T}$ is the signed throughput.
We note that purely specular materials are not modified, as they behave strictly as ideal mirror surfaces.

With both signed radiance and signed reflectance enabled, we can effectively model the observed behavior that latents can take on values outside of the standard rendering range while still maintaining the structure of the shading and lighting effects present in the original image.  Our subsequent modifications use this signed rendering as a starting point, and aim to model the rendered latents in a way that will capture features that deviate from a physical basis.

\subsubsection{Flat response term}

Our latent analysis indicates that, in some channels, different surfaces exhibit a comparatively flat appearance rather than faithfully reflecting their original material properties. To account for this behavior, we introduce an additional flat response term applied at surface interactions, which provides explicit control over the extent to which objects preserve material-dependent variation versus collapsing to a near-constant response. The flat response term $\tilde{L}_a$ is a signed radiance value, and its contribution can be estimated with:

\begin{align} \label{eq:ambient}
\tilde{I}_i^a(\sceneparams)=\int_{\pathspace_{E}}\pixelfilter_i(\xpath) {\tilde{T}}(\xpath,\sceneparams)\tilde{L}_a(\xp_d, \pi)d\xpath \,.
\end{align}
where the path space $\pathspace_{E}$ denotes a camera subpath $\xpath$ following the $ES^*D$ format, i.e., a path starting at the camera sensor ($E$) with an optional series of specular reflections ($S^*$) followed by a diffuse interaction ($D$).
Note that it is valid to define the flat-response term for full light transport paths, by accumulate it at each interaction, following the original path space $\pathspace$; however, we empirically observe that this does not lead to a significant performance difference, and hence opt for the simpler design using only camera subpaths. %

\subsubsection{Occlusion term} Our latent analysis also shows that shaded regions can deviate from the physically based interpretation of light transport, exhibiting higher intensity than unoccluded regions of the same material, or even values of opposite sign. To model this behavior, we introduce an occlusion term that acts as an additional contribution emitted only from shaded regions. The occlusion term $\tilde{L}_c$ stores signed radiance values, and its contribution to the rendered intensity is estimated as

\begin{align} \label{eq:occ}
\tilde{I}_i^c(\sceneparams)=\int_{\pathspace_{E}}\pixelfilter_i(\xpath) \tilde{T}(\xpath,\sceneparams)\tilde{L}_{c}(\xp_d, \pi)V(\xp_{d}, \yp)d\xpath \,.
\end{align}

Here, we leverage camera subpaths $\xpath \in \pathspace_E$ with an ending diffuse vertex $\xp_d$ to evaluate this integral. We sample an additional light vertex $\yp$ on a light source and evaluate the binary visibility function $V(\xp_{d},\yp)$, which equals one when the ending diffuse vertex $\xp_d$ is visible from $\yp$, and zero otherwise.

\subsubsection{Latent rendering equation} By combining the above modifications, we obtain our latent rendering equation:
\begin{align} \label{eq:latenteq}
    \tilde{I}_i(\sceneparams)=\tilde{I}_i^e(\sceneparams)+\tilde{I}_i^a(\sceneparams)+\tilde{I}_i^c(\sceneparams) \,.
\end{align}

To estimate this latent intensity, we construct a single light path $\xpath$ using BSDF sampling and evaluate all corresponding integrals along that path. $\tilde{I}^a$ and $\tilde{I}^c$ are estimated when we encounter the first diffuse surface, whereas we continue to extend the path to estimate $\tilde{I}^e$.
To efficiently estimating $\tilde{I}^e$, BSDF sampling is combined with next-event estimation (NEE) using multiple importance sampling, where the signed emitters ($\tilde{L}_e$) are explicitly sampled.

Note that rendering systems simulate lighting in linear radiance values, whereas generative models assume gamma-corrected inputs. For this reason, we apply gamma correction with $\gamma=2.2$ to the positive and negative values of the rendered latents, respectively:
\begin{align}
g(x, \gamma)=\max(x,0)^{\gamma}-\max(-x,0)^{\gamma} \,.
\end{align}

\subsection{Optimization}  
We use a differentiable renderer~\cite{Jakob:2022:DrJit} to perform latent rendering (\cref{eq:latenteq}). 
We optimize the scene parameters to make the renderer produce latents following the latent space defined by the VAE encoder. 
All latent channels are jointly rendered, analogously to RGB rendering. For a given scene and a reference latent map from a training view, we optimize the parameters $\pi = \{ \tilde{L}_e, \tilde{L}_a, \tilde{L}_c \}$ for each path vertex, which is represented as a vector in $\mathbb{R}^d$ with $d$ as the total latent channels.
We also optimize the reflectance parameters of each BSDF present in the scene, as well as the alpha parameters of any rough BSDFs. We support diffuse materials, conductors, dielectrics, and plastics and their rough variants.
We apply a clipping function to make sure the learnable parameters are non-negative during optimization.

To train the renderer, we minimize a Huber loss between the gamma-corrected rendered latents $\tilde{I}$ and ground-truth latents $I_\mathrm{gt}$: 
\begin{align}
    \mathcal{L}_{\pi} = 
    \begin{cases}
        \frac{1}{2} \| \tilde{I} - \tilde{I}_\mathrm{gt} \|_2^2 & \text{if } \| \tilde{I} - \tilde{I}_\mathrm{gt} \|_1 \leq \delta \\
        \delta \left( \| \tilde{I} - \tilde{I}_\mathrm{gt} \|_1 - \frac{1}{2} \delta \right) & \text{if }  \| \tilde{I} - \tilde{I}_\mathrm{gt} \|_1 > \delta
    \end{cases}
\end{align}
We set $\delta=1$, which can make optimization more stable when faced with large but infrequent per-pixel errors due to the strongly encoded edges observed in the pilot study.
Note that we compute the loss directly in the latent space, which empirically improves training convergence, stability, and efficiency. 
This also helps avoid inefficient gradient backpropagation through the decoder.

\subsection{Latent refinement}
\label{subsec:neuralrefiner}
The rendered latents produced by our fitted scene parameters and latent rendering equation impose a strong physically based prior.
However, as discussed in the latent analysis, latent values can deviate from physically based behavior, particularly around object boundaries and other high-frequency regions. 
We train a feed-forward neural network to predict these residual latents. 
For simplicity, we train this neural refiner on a per-scene basis using a single ground truth rendering. We leave the generalization of the refiner across scenes for future work.

The refiner network $R_\theta$ takes as input the rendered latent $\tilde{I}$ along with a set of auxiliary buffers (AOVs) denoted as $A$. These AOVs include surface shading normals and depth, providing geometric cues that guide the network toward correcting structure-dependent artifacts. The AOVs are rendered at the latent resolution and concatenated to the input in a channel-wise manner. 
The network outputs the residual latents as $R_\theta(\tilde{I}, A)$. 

We adopt a feed-forward multilayer perceptron architecture for the refiner network. The hidden layer width is chosen as a multiple of the number of latent channels $d$. For Stable Diffusion 3.5, we use an expansion factor of 16, resulting in hidden layers with $16 \times 16 = 256$ channels. 
The network then consists of three convolution blocks with kernel sizes $3 \times 3$, $5 \times 5$, and $3 \times 3$.

We train the refiner in a separate stage after the physically based renderer is optimized. 
To train the refiner, we use a composite loss defined in both the latent and RGB spaces: %
\begin{align}
\mathcal{L}_\theta = \MSE(\tilde{I}', \tilde{I}_\mathrm{gt}) + \SSIM(\tilde{I}', \tilde{I}_\mathrm{gt}) + \LPIPS(\dec(\tilde{I}'),\dec(\tilde{I}_\mathrm{gt})) \,,
\end{align}
where the refined latent is $\tilde{I}' = \tilde{I} + R_\theta(\tilde{I}, A)$ and $\dec$ denotes the decoder.
The MSE loss and SSIM loss enforces the refined latents to stay close to the ground-truth latents. The LPIPS loss~\cite{zhang2018perceptual} applies perceptual similarity constraints on the decoded image, reducing color shifts and encouraging sharper boundaries. 

\section{Experimental Results}

\subsection{Implementation details} 

We use Mitsuba~\cite{Jakob:2022:DrJit} with path replay backpropagation~\cite{Vicini2021PathReplay}, without reparameterization~\cite{Zhang2023Projective}, to perform the optimization. We extend Mitsuba to support signed rendering with 16 latent channels, enabling efficient joint optimization over all channels. We also support textured materials: each RGB reflectance texture is converted to grayscale and then mapped into the 16-channel latent space via a learnable per-channel scale and offset. This preserves the texture's spatial detail while exposing its latent appearance as optimizable parameters. Training follows a multi-stage procedure: we first optimize the scene parameters for 3000 iterations with a learning rate of $10^{-4}$, and then train the refinement network for an additional 3000 iterations using the same learning rate. All optimizations use the Adam optimizer \cite{kingma2014adam} with default hyperparameters. We plan to release our implementation upon acceptance.

To improve stability during scene parameter optimization, we employ a learning-rate scheduler following a cosine annealing schedule with restarts, as well as early stopping. On a simple scene such as the Cornell Box, training on a single NVIDIA RTX~4090 GPU takes 5 minutes for scene optimization and 5 minutes for refiner training. Scene parameters $\sceneparams$ are initialized randomly from a uniform distribution in $[0,1]$ for both positive and negative components of all trainable parameters, as well as the BSDF parameters. 

We compare our method with a baseline that uses the traditional rendering formula~\cite{Kajiya:1986:Rendering} for latent rendering. We produce 16-channel values in the $[0,1)$ range with the traditional rendering process, then de-normalize the rendered latents back into the native latent range $\sim[-5,5]$ using the logit function $\log(x/(1-x))$, and train the refiner on these de-normalized latents following~\cref{subsec:neuralrefiner}. The refiner output is passed directly to the decoder. Compared to our method, this baseline skips the latent modeling in~\cref{subsec:latentrender} and simply uses the traditional rendering formula.

\subsection{Scene setups}
We analyze the performance of our latent rendering method on five scenes of increasing complexity: Cornell Box, Lamp, Veach-Bidir, Dining Room, and Living Room. The Cornell Box scene contains a single light source and a mix of diffuse and pure specular materials. The Lamp scene includes multiple light sources and rough plastic materials. The Veach-Bidir scene contains difficult-to-sample light sources and caustics effects. The Dining Room and Living Room scene features a single light source and multiple objects with distinct BSDFs. For Cornell Box and Veach-Bidir, we associate each object with a separate BSDF with its own learnable parameters. For Lamp, Dining Room, Living Room scenes, we use the default shape-BSDF associations provided in original scenes to reduce memory consumption during the optimization of the BSDF parameters.
We assign each object with a non-delta BSDF its own flat-response and occlusion parameters, $\tilde{L}_a$ and $\tilde{L}_c$, which have random initializations. 
For each scene, we provide reference latents obtained from a single view. These reference latents are obtained by encoding the RGB rendering produced with path tracing on large number of samples per pixels (e.g., 16K spps). %

\begin{figure}[t]
    \centering
    \includegraphics[width=\linewidth]{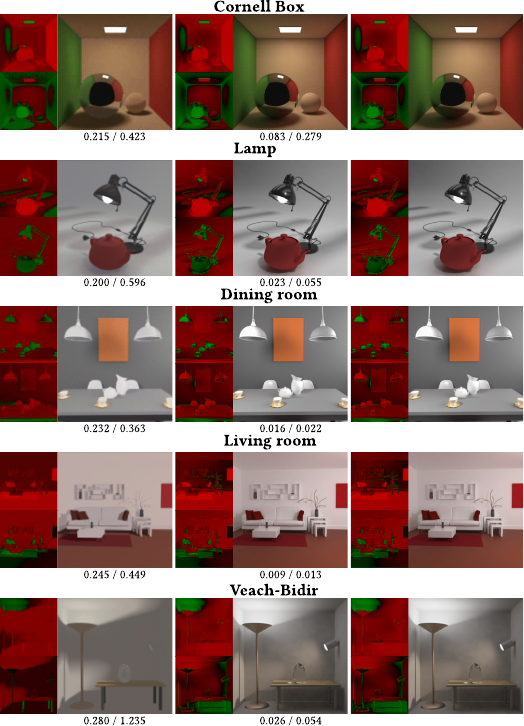}
    \caption{We compare the performance of the training view on rendered latent and refined latent with the ground truth. Representative latents for each case are shown on the left of each color rendering. Metrics shown are LPIPS/MSEx100.
    }
    \label{fig:matrix-training}
\end{figure}

\begin{figure}[t]
    \centering
    \includegraphics[width=\linewidth]{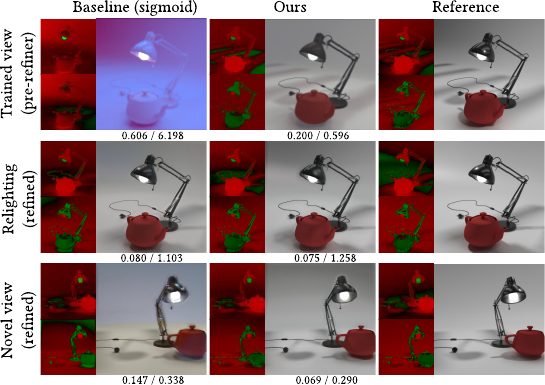}
    \vspace{-4mm}
    \caption{%
    Comparison between the baseline and our method on the Lamp scene. We observe that our latent modeling and rendering formula provide significant quality improvement compared to the baseline that only relies on latent normalization and traditional rendering. 
    Metrics shown are LPIPS/MSEx100.
    }
    \label{fig:baseline}
\end{figure}

\begin{figure}[t]
    \centering
    \includegraphics[width=\linewidth]{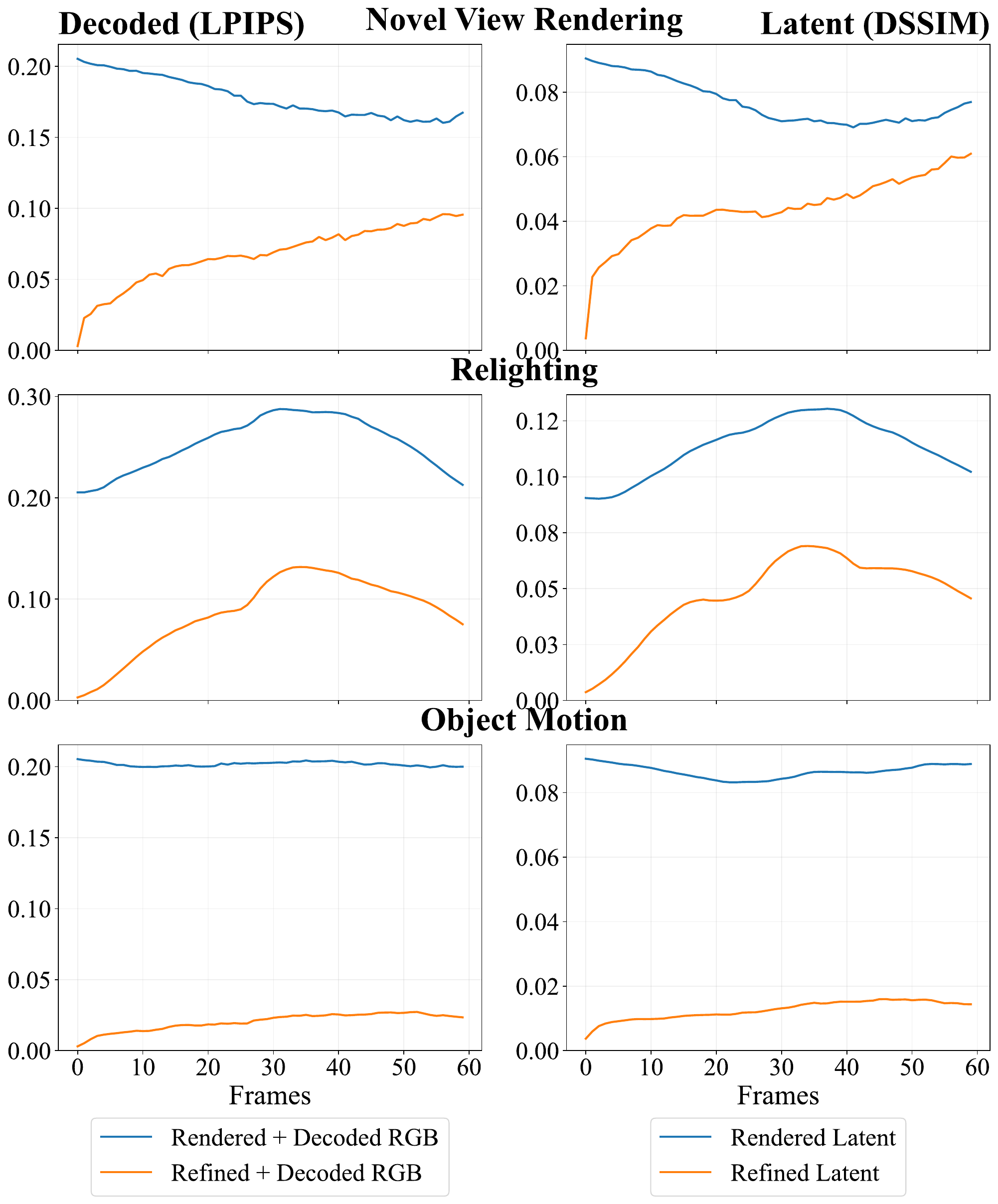}
    \vspace{-4mm}
    \caption{The error of the Lamp scene across 60 frames of scene editing variants. Frame 0 corresponds to the training view. %
    }
    \label{fig:error-anim}
\end{figure}

\begin{figure}[t]
    \centering
    \includegraphics[width=\linewidth]{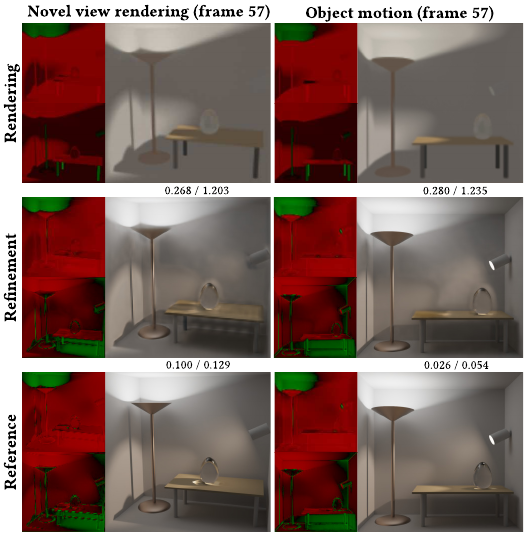}
    \caption{Qualitative results on the Veach-Bidir scene. Metrics shown are LPIPS/MSEx100. %
    }
    \label{fig:matrix-bidir}
\end{figure}

We demonstrate the capabilities of our latent rendering framework on scene editing tasks. In all experiments, the scene geometry, material types (e.g., diffuse or specular), emitter locations, and camera pose are assumed to be known. 
We perform scene edits and render latents in the edited scene to explore the generalization of latent rendering to various scenarios. 
We consider camera movement for novel view synthesis, light source movement for scene relighting, and object movement for animated rendering. 
Relaxing the scene input assumptions such as taking a single image as input and predicting scene geometry and PBR materials~\cite{wang2025moge,xiang2025trellis2} would be an interesting direction for future work. 
We compare the latents and their reconstruction with ground truth latents and images, respectively. 
We report comparison results using the MSE, SSIM~\cite{wang2004image}, and LPIPS metrics.

\subsection{Qualitative results} 

\cref{fig:matrix-training} demonstrates how our method performs before and after refinement on the training view. For completeness, we show two sets of representative latents for each decoded RGB image. We observe, across all scenes, that the main content of the latents and their decoded images is produced by the latent renderer, as evidenced by their high similarity to the ground truth. The refiner is then able to overfit and deliver an almost perfect reconstruction.

\begin{figure*}
    \centering
    \includegraphics[width=\linewidth]{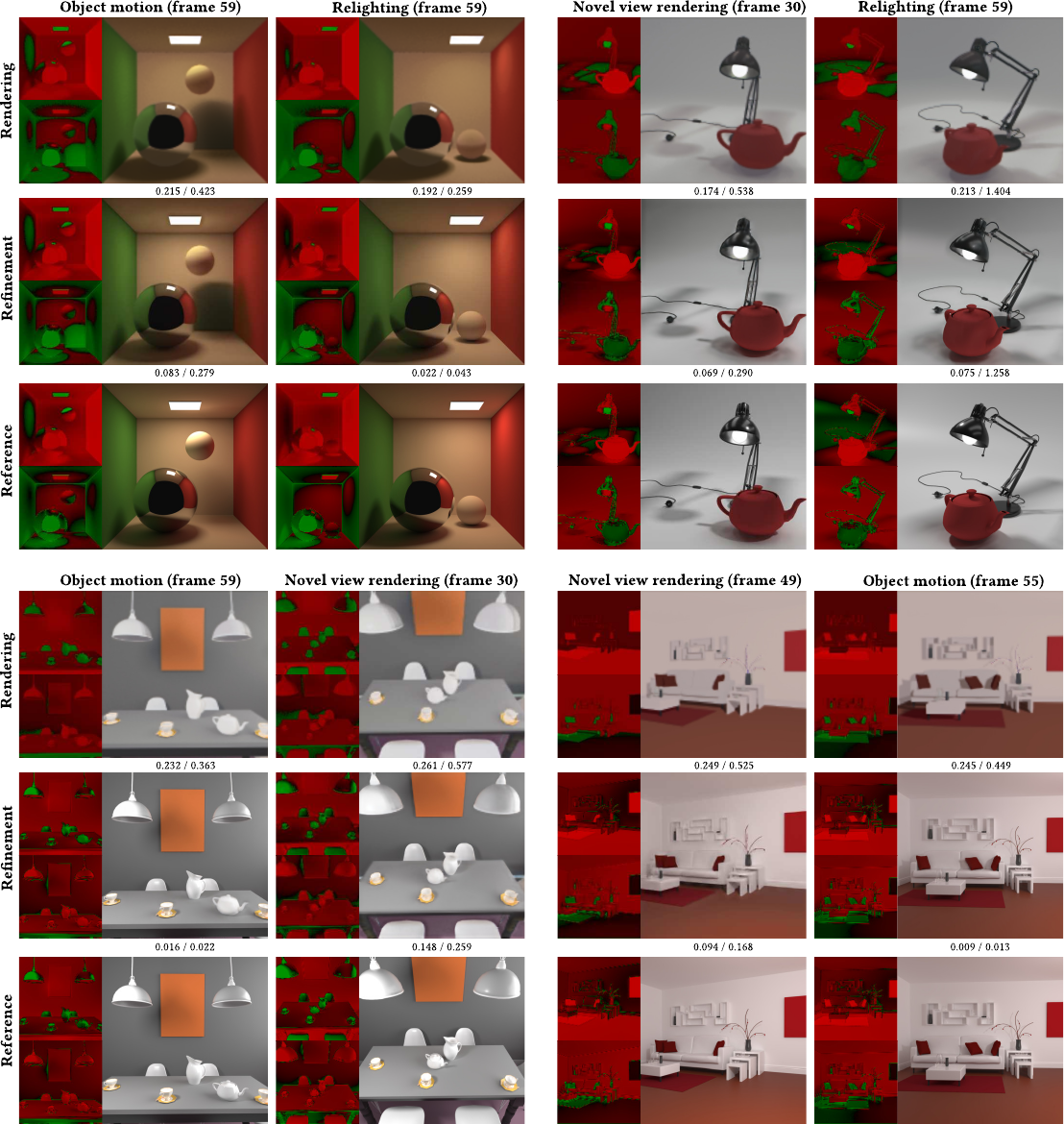}
    \caption{We compare the performance of the scene editing on rendered latent and refined latent with the ground truth. Representative latents for each case are shown on the left of each color rendering. Metrics shown are LPIPS/MSEx100. 
    }
    \label{fig:matrix-views}
\end{figure*}

\cref{fig:matrix-views,fig:matrix-bidir} provide the rendering results when the scene is edited in various scenarios. We can observe that our method generalizes well to novel configurations: most of the work is done by the rendered latents, while the refiner is responsible for improving object boundaries and increasing overall image contrast and sharpness. 

For the training view, we found that the post-refinement reconstruction is almost indistinguishable from the ground truth image, which confirms the success of the training. 
When the scene is edited, through camera motion, changes in light position, or object movement, our method can regenerate images that remain visually close to the ground truth. 
This result demonstrates two advantages of latent rendering: the trained renderer can be applied to synthesize new latents that can be reconstructed to meaningful images, and the refiner generalizes to unseen settings. 
Notably, specular reflections and shadowing effects can be reconstructed based on the visual cues in the physically based rendered latents. 

\begin{figure*}
    \centering
    \includegraphics[width=\linewidth]{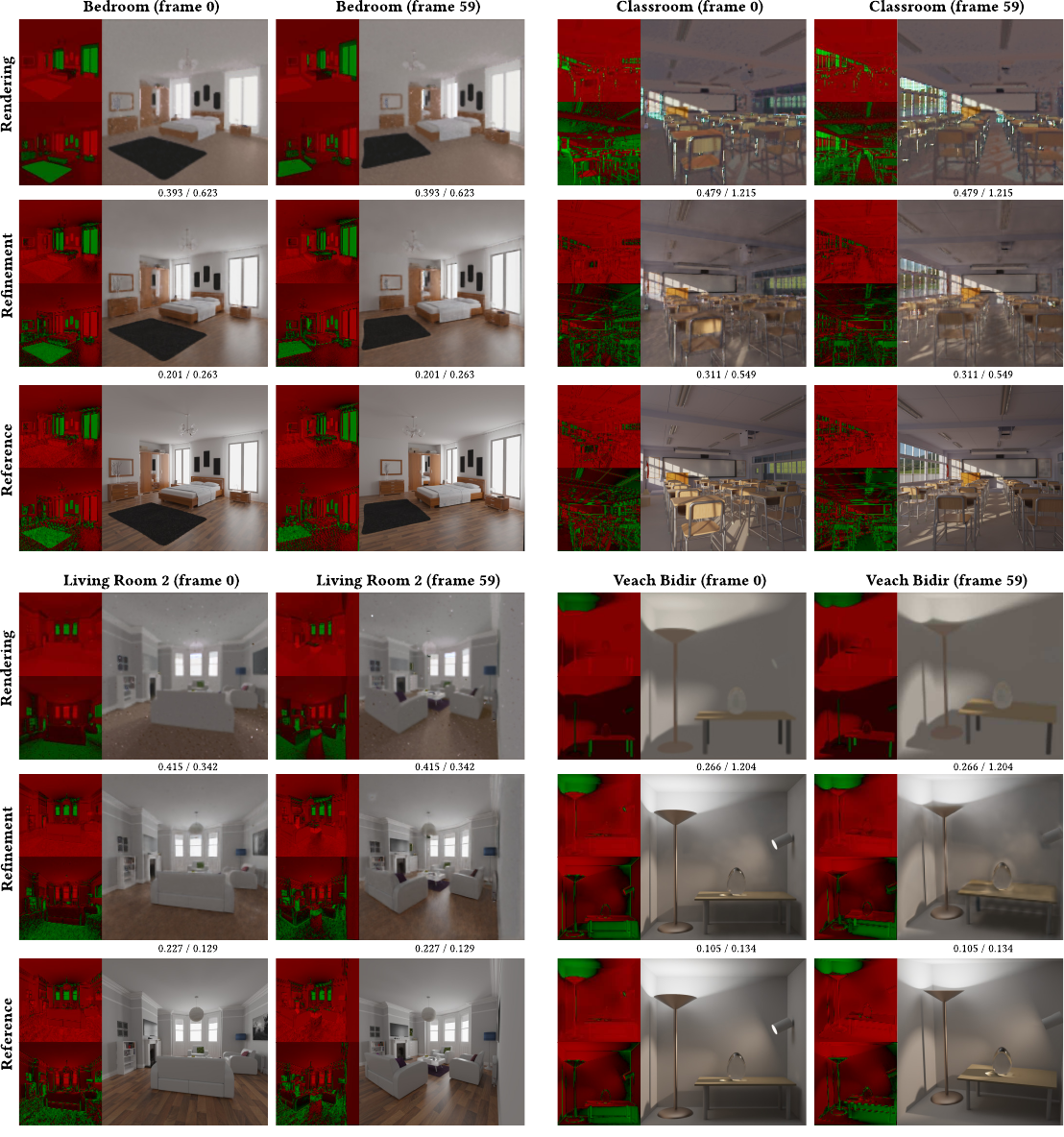}
    \caption{
        We compare the performance of novel view scene editing on rendered latents and refined latents with the ground truth.  We showcase complex scenes with texture support. Metrics shown are LPIPS/MSEx100.
    }
    \label{fig:matrix-textured}
\end{figure*}

\cref{fig:matrix-textured} shows novel view rendering on three additional scenes (Bedroom, Classroom, and Living Room 2) with texture support. It shows that latent rendering with texture is promising, with further optimizations for quality improvement left as future work.
Video results for each scene are provided in the supplemental material.

\paragraph*{Baseline comparison.} We provide baseline comparisons in~\cref{fig:baseline}. Compared to our method, the rendered latents of the baseline yield color shifts in the reconstruction (1st row). This could be explained by the non-linear range compression and expansion due to normalization and de-normalization, respectively.
The refiner significantly improves the rendered latents and the final rendering, but fails to achieve accurate color rendering while having more visual artifacts in the baseline (2nd and 3rd rows).
This result confirms the necessity and effectiveness of our latent rendering equation.

\paragraph*{Errors of scene edits.} \cref{fig:error-anim} provides the error plot of the scene edits for the Lamp scene. The edits are performed for 60 frames, starting from frame 0 as the training view. 
We use D-SSIM to measure structural difference and LPIPS to measure perceptual difference to the ground truth latents.  
For all tasks, the rendered latents generalize fairly well across frames with similar errors except for the extreme case in the light movement (around frame 30). The refiner improves the rendered latents significantly across frames. As the refiner is only trained on a single view, D-SSIM and LPIPS are smaller (i.e., better) at views more similar to the training view.

\subsection{Quantitative results}

\begin{table}[t]
\centering
\small
\setlength{\tabcolsep}{3pt}
\renewcommand{\arraystretch}{1.15}
\caption{Ablation results reporting the mean LPIPS / MSE$\times 100$ achieved over the decoded RGBs. %
``Var.'' denotes the scene-edit variant: camera, object, and light motion.}

\label{tab:mean}
\begin{tabular}{@{}cl cccc@{}}
\toprule
& {Var.} & $\tilde{I}_i^e$ (\cref{eq:signedrender}) & $+\,\tilde{I}_i^a$ (\cref{eq:ambient}) & $+\,\tilde{I}_i^c$ (\cref{eq:occ}) & {+\,Refiner} \\
\midrule
\multirow{3}{*}{\rotatebox{90}{\makecell{Cornell\\Box}}} & Cam. & 0.502 / 2.264 & 0.228 / 0.335 & 0.213 / 0.222 & 0.075 / 0.069 \\
 & Obj. & 0.526 / 2.355 & 0.213 / 0.338 & 0.202 / 0.261 & 0.032 / 0.061 \\
 & Light & 0.519 / 2.361 & 0.209 / 0.330 & 0.191 / 0.220 & 0.013 / 0.024 \\
\midrule
\multirow{3}{*}{\rotatebox{90}{Lamp}} & Cam. & 0.394 / 3.091 & 0.206 / 0.844 & 0.178 / 0.564 & 0.068 / 0.297 \\
 & Obj. & 0.439 / 3.274 & 0.236 / 0.791 & 0.202 / 0.608 & 0.020 / 0.048 \\
 & Light & 0.504 / 5.702 & 0.292 / 3.000 & 0.253 / 2.385 & 0.086 / 1.867 \\
\midrule
\multirow{3}{*}{\rotatebox{90}{\makecell{Living\\Room}}} & Cam. & 0.643 / 5.502 & 0.281 / 0.599 & 0.250 / 0.480 & 0.072 / 0.115 \\
 & Obj. & 0.648 / 5.584 & 0.280 / 0.573 & 0.246 / 0.451 & 0.008 / 0.013 \\
 & Light & 0.659 / 6.031 & 0.299 / 1.053 & 0.266 / 0.894 & 0.064 / 0.392 \\
\bottomrule
\end{tabular}
\vspace{-2mm}
\end{table}

\paragraph*{Ablation studies.} \Cref{tab:mean} presents our ablation study of the latent formulation and the impact of the refiner on the final decoded image across different scene edits. In this table, we report the mean LPIPS and MSE values attained along each animated editing sequence (moving camera, object and emitter). Using only signed emitters (\cref{eq:signedrender}) leads to poor performance and unstable training. In contrast, the flat-response term (\cref{eq:ambient}) and occlusion term (\cref{eq:occ}) have a significant impact on image quality. The neural refiner further improves the latents, enabling higher quality image reconstruction. 
Note that the refiner only acts on predicting the residual latents once the rendered latents are available. If the rendered latents are noisy (e.g., produced using only signed emitters), the refiner does not yield significant improvement. 

\begin{figure}
    \centering
    \includegraphics[width=0.95\linewidth]{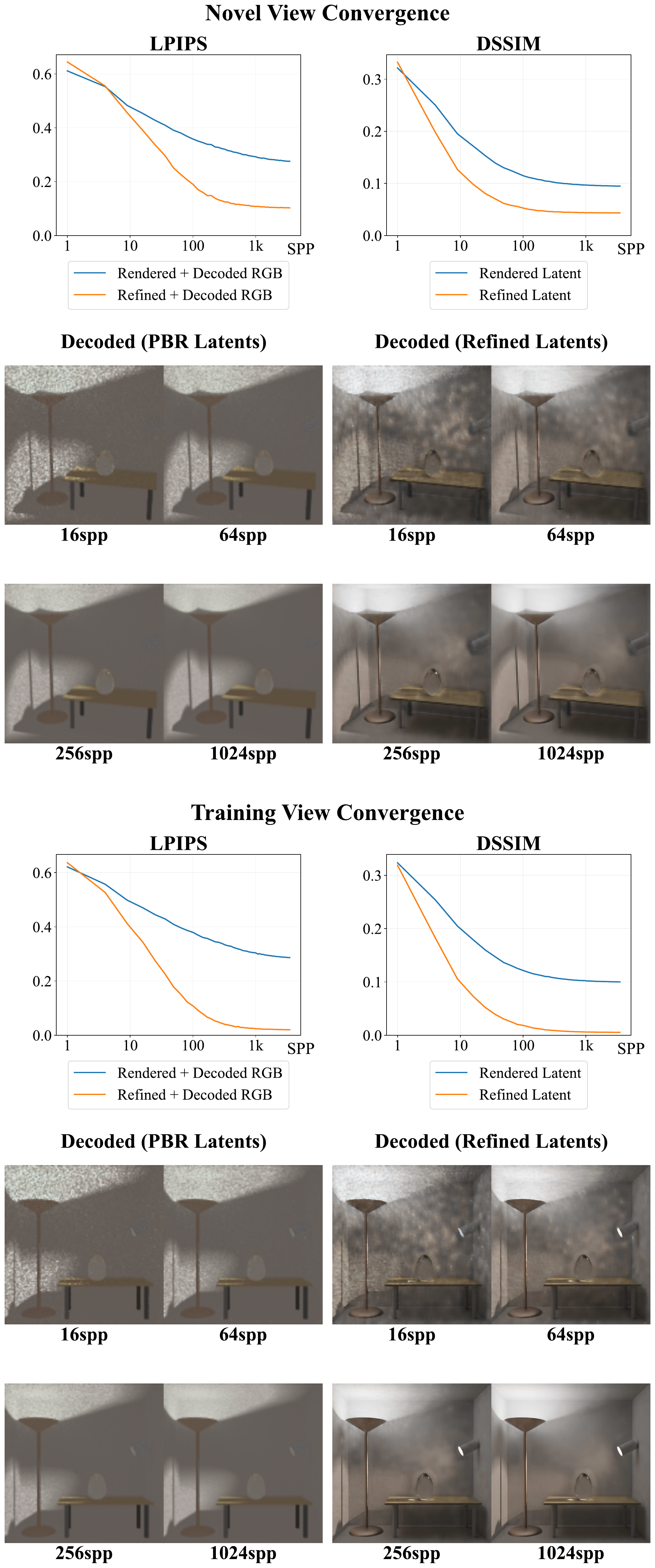}
    \vspace{-2mm}
    \caption{Empirical convergence vs.\ samples per pixel (spp) for a novel view (top) and a training view (bottom) of the Veach-Bidir scene. LPIPS on the decoded RGB (left) and DSSIM on the latents (right).}
    \label{fig:convergence}
\end{figure}

\paragraph*{Convergence analysis.} Theoretically, latent rendering follows the Monte Carlo estimation framework, and hence, the convergence properties in the latent space remains the same as in the RGB space. 
We use path tracing in this work, and therefore, the estimation of the rendered latents are unbiased. 
However, the use of the refiner network results in biased latents, and the image reconstruction using the decoder results in biased final rendering. 
Ideally, the rendering would become unbiased if the refiner and the decoder were linear functions. 
Empirically, we provide convergence plots of our rendering in~\cref{fig:convergence}. 
We observe that the convergence patterns are similar across scenes, and therefore provide the convergence of the Veach-Bidir scene as a representative. 
The plots confirm that, for the training view, both the latent and the decoded RGB estimates converge as the number of samples per pixel increases. The novel view exhibits a similar trend, with a small residual bias introduced by the refiner. %

\subsection{Runtime analysis}

\begin{figure*}
    \centering
    \includegraphics[width=0.95\linewidth]{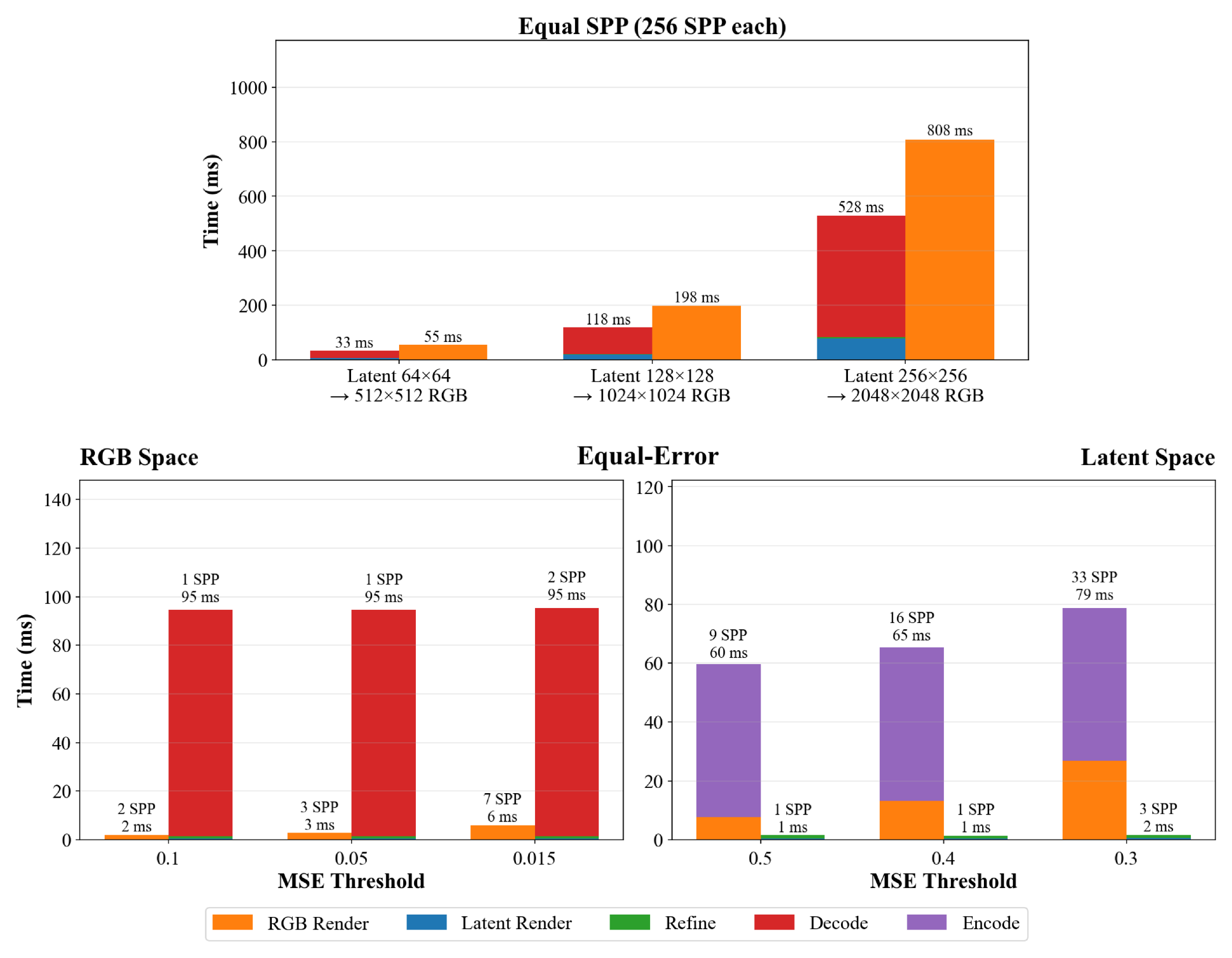}
    \vspace{-2mm}
    \caption{Equal-error runtime comparison between RGB path tracing and our
latent rendering method (with refiner), for RGB space (left) and latent space (right). The plot is based on frame 30 of the move\_camera test case of the Lamp scene.}
    \label{fig:equal-error}
\end{figure*}

We evaluate the runtime of latent rendering against RGB path tracing in an \emph{equal-error} setting: for each method we measure the time and the number of samples per pixel required to reach a fixed MSE threshold (\cref{fig:equal-error}).

When the goal is to estimate values in the \emph{latent space} (right), our method is significantly faster ($42$--$84\times$) than first path tracing in RGB and then encoding, since the latent render and refiner are efficient and requires less samples. When the goal is a decoded \emph{RGB image} (left), however, our method is not competitive with direct RGB path tracing: despite the spatial compression, the single decoder evaluation dominates the runtime (the red bar) and makes our method $6$--$35\times$ slower. Accelerating the decoder 
is therefore key to making latent rendering competitive for RGB output, and is an interesting direction for future work.

\subsection{Applications}

\begin{figure*}[t]
    \centering
    \begin{minipage}{0.3\linewidth}
    \includegraphics[width=\linewidth]{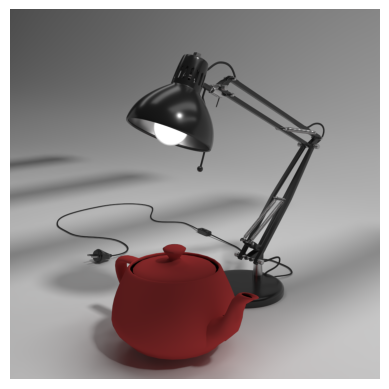}\\
    \centerline{(a) Original scene}
    \end{minipage}
    \begin{minipage}{0.3\linewidth}
    \includegraphics[width=\linewidth]{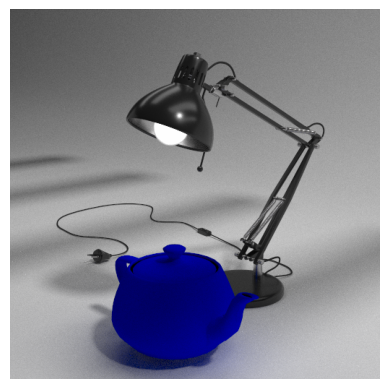}\\
    \centerline{(b) SDS with RGB rendering + encoder}
    \end{minipage}
    \begin{minipage}{0.3\linewidth}
    \includegraphics[width=\linewidth]{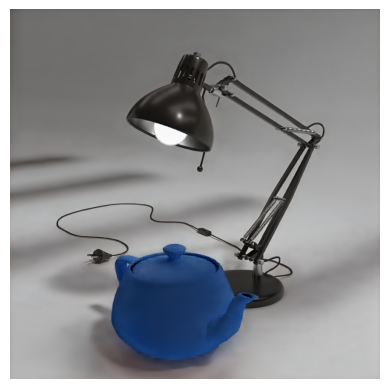}\\
    \centerline{(c) SDS with latent rendering} 
    \end{minipage}
    \caption{Application using score distillation sampling (SDS)~\cite{poole2023dreamfusion}.
    We use the pretrained diffusion model to iteratively update the reflectance of the teapot following guidance from the text prompt ``A \textcolor{blue}{blue} teapot and a lamp on a table.''}
    \label{fig:sds}
\end{figure*}

\noindent\textbf{Scene editing.}
Latent rendering could be further explored for downstream applications. For example, as latent rendering naturally operates in the latent space, one can integrate score distillation sampling with latent rendering, enabling the optimization of scene parameters guided by a pre-trained diffusion prior~\cite{poole2023dreamfusion}. 
Image editing applications could be integrated with latent rendering as well, e.g., object insertion and background removal using blended latents~\cite{avrahami2023blendedlatent}, image harmonization and semantic image composition using delta denoising distillation~\cite{hertz2023delta,chen2024freecompose}.

We provide an example of using score distillation~\cite{poole2023dreamfusion} to re-color the teapot in the Lamp scene by optimizing its BSDF reflectance. The editing is guided by the text prompt ``a blue teapot and a lamp on a table''.
Traditional SDS operates the optimization in the latent space, and therefore it requires first performing the rendering in the RGB space, and then encode the rendered image using the encoder. 
By using latent rendering, we can directly compute the SDS objective based on the refined latents, skipping the encoder in the optimization pipeline.
The results are shown in \cref{fig:sds}.
Both the RGB rendering approach and our latent rendering approach can successfully update the teapot color. Due to the per-scene training of the refiner, our teapot results in some blurs that could be further corrected by improving the generalization of the refiner.

\subsection{Limitations and future work}
Our method is a very first step toward latent rendering. 
Several limitations remain for further investigations in future work. 
First, as the latent space is in low spatial resolutions, aliasing tends to dominate when rendering highly detailed objects. 
Second, while our refiner does fairly well at generalizing to intra-scene changes, it is trained only on a single image and with a per-scene setting, and thus does not generalize across scenes. 
Building a universal refiner across scenes would improve the practicality of the method.
Finally, while latent rendering offers a new angle for extending physically based rendering research, there are still rooms for improving image quality and running time to match the performance of traditional methods. 

As future work, we aim to explore latent rendering with state-of-the-art visual diffusion models and pursue more accurate latent representations and processing for extended applicability and improved computational efficiency.  
For example, while our trained renderer and latent refiner generalizes well to higher resolution rendering, it is possible to add neural upsampling. Latent space super resolution~\cite{jeong2025latent} validated that 
since latent space represents image features in a spatially compressed format with high-level information, performing upsampling in this domain preserves well details and sharpness and could generate more details compared to RGB upsampling. Applying this method to improve our latent rendering would be future work.

\section{Conclusion}

We have presented latent rendering, a paradigm that trains a physically based renderer to output latent representations defined by the autoencoders of a generative model. The rendered latents can be reconstructed to color images following a neural refiner and an image decoder.
Applications include direct editing of a 3D scene rendering in latent space, enabling a straightforward process to render scene variants under camera, lighting, and object changes. 
We believe that latent rendering is a promising approach to unlock the connection between physically based rendering and generative models. By leveraging latent spaces, we can enable the integration of physical priors directly into creative content generation.

\paragraph*{Acknowledgment.} We would like to thank
Wig42 (\textsc{Dining Room} and \textsc{Living Room}),
Jay-Artist (\textsc{Living Room 2}),
NovaZeeke (\textsc{Classroom}),
SlykDrako (\textsc{Bedroom}),
UP3D (\textsc{Lamp}),
and Benedikt Bitterli~\cite{resources16}
for providing the scenes, which we obtained from the Mitsuba~3 gallery.
This project is supported by Research Ireland under the Research Ireland Frontiers for the Future Programme - Project, award number 22/FFP-P/11522.

\bibliographystyle{eg-alpha-doi.bst}
\bibliography{references}

\end{document}